\pdfoutput=1
\documentclass[%
 reprint,showkeys,
superscriptaddress,
 amsmath,amssymb,
 aps,
floatfix,
]{revtex4-1}

\usepackage{xcolor}
\usepackage{graphicx}
\usepackage{dcolumn}
\usepackage{bm}
\usepackage{hyperref}
\usepackage{natbib}
\usepackage[utf8]{inputenc} 

\begin{document}

\preprint{APS/123-QED}
\title{Dynamic local strain in graphene generated by surface acoustic waves}

\author{Rajveer Fandan}
\email{rajveer.fandan@upm.es}
\author{Jorge Pedrós}
\email{j.pedros@upm.es}
\affiliation{Instituto de Sistemas Optoelectrónicos y Microtecnología, Universidad Politécnica de Madrid, Av. Complutense 30, Madrid 28040, Spain}
\affiliation{Departamento de Ingeniería Electrónica, E.T.S.I de Telecomunicación, Universidad Politécnica de Madrid, Av. Complutense 30, Madrid 28040, Spain}
\author{Alberto Hernández-Mínguez}
\affiliation{Paul-Drude-Institut für Festkörperelektronik, Leibniz-Institut im Forschungsverbund Berlin e.V., Hausvogteiplatz 5-7, 10117 Berlin, Germany}
\author{Fernando Iikawa}
\affiliation{Paul-Drude-Institut für Festkörperelektronik, Leibniz-Institut im Forschungsverbund Berlin e.V., Hausvogteiplatz 5-7, 10117 Berlin, Germany}
\affiliation{Institute of Physics, State University of Campinas, 13083-859 Campinas-SP, Brazil}
\author{Paulo V. Santos}
\affiliation{Paul-Drude-Institut für Festkörperelektronik, Leibniz-Institut im Forschungsverbund Berlin e.V., Hausvogteiplatz 5-7, 10117 Berlin, Germany}
\author{Alberto Boscá}
\author{Fernando Calle}
\affiliation{Instituto de Sistemas Optoelectrónicos y Microtecnología, Universidad Politécnica de Madrid, Av. Complutense 30, Madrid 28040, Spain}
\affiliation{Departamento de Ingeniería Electrónica, E.T.S.I de Telecomunicación, Universidad Politécnica de Madrid, Av. Complutense 30, Madrid 28040, Spain}

\date{\today}

\begin{abstract}
We experimentally demonstrate that the Raman active optical phonon modes of single layer graphene can be modulated by the dynamic local strain created by surface acoustic waves (SAWs). In particular, the dynamic strain field of the SAW is shown to induce a Raman scattering intensity variation as large as $15\:\%$ and a phonon frequency shift of up to 10 cm$^{-1}$ for the G band, for instance, for an effective hydrostatic strain of $0.24\:\%$ generated in a single layer graphene atop a LiNbO$_{3}$ piezoelectric substrate with a SAW resonator operating at a frequency of $ \sim $ 400 MHz. Thus, we demonstrate that SAWs are powerful tools to modulate the optical and vibrational properties of supported graphene by means of the high-frequency localized deformations tailored by the acoustic transducers, which can also be extended to other 2D systems. 
\end{abstract}

\maketitle

\section{Introduction}

Graphene is a well-known 2D material with extreme properties, including a room temperature mobility of up to $10^{5}$ cm$^{2}$V$^{-1}$s$^{-1}$ \cite{Mayorov2011}, a thermal conductivity of 3000 Wm$^{-1}$K$^{-1}$ \cite{Balandin2011}, a Young’s modulus of 1 TPa, an intrinsic strength of 130 GPa, and the capability of sustaining an elastic tensile strain of up to $25\:\%$ \cite{Lee2008}. Moreover, strain in graphene has been shown to give rise to various extraordinary phenomena such as the shifting of the Dirac cones \cite{Pereira2009}, the shifting and splitting of the graphene Raman modes \cite{Huang2009,Yoon2011}, the enhancement of the electron–phonon coupling and superconductivity \cite{Si2013}, the generation of pseudomagnetic fields \cite{Levy2010}, and the zero-field quantum Hall effect \cite{Guinea2009}. In addition, strain has also been reported to affect the interaction of the graphene surface with the environment, leading, for example, to the stabilization of the adsorption of metal atoms by preventing their clustering \cite{Zhou2010,Cretu2010}, or to the increase in the coverage and self-assembly of hydrogen atoms \cite{Ng2010,Wang2011}. Typically, strain is introduced by placing graphene on a stretchable and bendable substrate \cite{Huang2009,Mohiuddin2009,Frank2010,Yoon2011}. 
However, approaches capable of generating strain locally with fast actuation mechanisms are highly desirable for the development of integrated devices. Local strain can be attained in suspended graphene, either pressurized \cite{Lee2012, Metten2014, Shin2016} or actuated in nano- \cite{Chen2009, Bao2012} and microelectromechanical \cite{Goldsche2018} systems (NEMS and MEMS), and in graphene/polymer membranes where the polymer is actuated by e-beam irradiation \cite{Polyzos2015,Colangelo2018}. Other local strain techniques require transferring the graphene onto nano-patterned substrates  \cite{Mi2015, Jiang2017} or nano-indentation \cite{Nemes-Incze2017}. However, the complexity of most of these methods hinders the usage of graphene in practical device configurations, where the effects induced by the locally generated strain could be exploited.  One convenient way to generate strain locally on supported graphene (or any other 2D material) is by means of a surface acoustic wave (SAW). The strain field of a SAW can be actively controlled allowing to change the magnitude of the dynamic strain electrically. This mechanism has already been proposed for creating an optical grating to launch surface plasmon polaritons in graphene \cite{Schiefele2013} and graphene/h-BN systems \cite{Fandan2018} and has been shown to modulate the emission from defect centres in h-BN \cite{Iikawa2019}. The strain from a sound wave has also been suggested to lead to collimation effect of the electron conduction \cite{Oliva-Leyva2015}. Additionally, the electric field accompanying a SAW in a piezoelectric substrate has also been proven useful for developing acoustoelectric devices in graphene \cite{Miseikis2012,Bandhu2013,Santos2013,Hernandez-Minguez2016,Liou2017} and MoS$_{\text{2}}$ \cite{Preciado2015}, as well as to modulate the excitonic response of the latter material \cite{Rezk2015}. Moreover, the piezoelectricity of few-layer MoS$_{\text{2}}$ with an odd-number of layers has been shown to modulate its excitonic response when a SAW travels across the material \cite{Rezk2016}. Therefore, SAWs might open new possibilites for strain engineering and straintronics in graphene and more generally in 2D materials \cite{Delsing2019}. 
     
In this work, we experimentally demonstrate the  modulation of the graphene phonon dispersion by means of the strain field of a SAW. In particular, we show that the G (optical phonon) and 2D (two optical phonons) Raman bands shift due to the phonon mode softening (hardening) under the tensile (compressive) strain of the SAW. The effect of a SAW on the modulation of optical phonons was earlier studied in bulk Si and GaN crystals \cite{Iikawa2016}. However, the achievable modulation in graphene is much stronger due to the larger Grüneisen parameters of graphene and the more efficient generation of the strain in the 2D layer on a strong piezoelectric substrate. Moreover, the whole 2D crystal lattice is coherently affected by the SAW strain field. We prove that this modulation can be actively controlled through the frequency and amplitude of the SAW generated by the RF signal applied to a piezoelectric transducer. Thus, the mechanical modulation generated by a piezoelectric substrate on the 2D systems can be a powerful tool to investigate their optical and vibrational properties and to develop new devices and applications. 

\section{Experimental Details}

A two-port SAW resonator (SAWR), formed by interdigital transducers (IDTs) and metal reflection gratings, was patterned on $128^{o}$ rotated Y-cut X-propagating LiNbO$_{\text{3}}$  by optical lithography followed by lift-off metallization. Single layer graphene (SLG), grown by chemical vapour deposition (CVD), was then transferred to the LiNbO$_{\text{3}}$ substrate using an automatic transfer system \cite{Bosca2016}. The graphene sheet was patterned using an oxygen plasma, leaving a graphene stripe in the region between IDTs, as shown in figure 1(a). The period and aperture of the IDTs were chosen to be $\lambda_{SAW} =$ 10 $\mu $m and $W_{IDT} =$ 100 $\mu $m, respectively. This $\lambda_{SAW}$ value permits to assess the SAW-induced effects by optical microspectroscopy, whereas the selected $W_{IDT} $ value is a trade-off providing large SAW power density, negligible diffraction effects in the SAW beam, and good impedance matching. Figure 1(b) shows the power reflection and transmission S parameters of the SAWR, measured with the device already mounted and wire-bonded on the chip used later on for the Raman characterization. 

The Raman spectroscopy was performed in backscattering configuration at room temperature using a confocal Horiba LabRam system. The incident 473-nm solid-state laser beam was focused onto a spot of approximately 1 $ \mu $m diameter and 0.8 mW power on the sample surface using a long working distance (50x) objective with a numerical aperture of 0.55. The backscattered light was collected by the same objective, spectrally analyzed by a monochromator with a 1800 grooves/mm dispersion grating and detected using a liquid-nitrogen-cooled charge coupled device (CCD). In order to detect the changes in the Raman spectra induced by the SAW, the amplitude-modulation method described in ref. \cite{Iikawa2016} was used. Here, both the RF signal generating the SAW at the IDTs and the excitation laser are chopped with the same modulation frequency (300 Hz) to be either in- or out-of-phase. Spectra were then recorded with the two excitations in phase (i.e. with the sample subjected to both the laser and acoustic excitations; called hereafter \textit{SAW ON}) and out-of-phase (where the sample is exposed only to the laser; called hereafter \textit{SAW OFF}) following the  alternated sequence \textit{SAW OFF} - \textit{SAW ON} - \textit{SAW ON} - \textit{SAW OFF}. An acquisition time of 60 s per spectrum was used for each excitation condition and 20 sequences were measured (80 spectra in total). By averaging, respectively, over the 40 \textit{SAW ON} and 40 \textit{SAW OFF} cycles, the Raman spectra with ($I_{on}$) and without ($I_{off}$) the effects of the SAW  were obtained. This method, besides increasing the signal-to-noise ratio due to the integration, minimizes the effects of the systematic fluctuation of the temperature and of the laser power on the Raman signal, allowing us to isolate the strain contribution \cite{Iikawa2016}. 

\begin{figure}
\includegraphics[width=\linewidth]{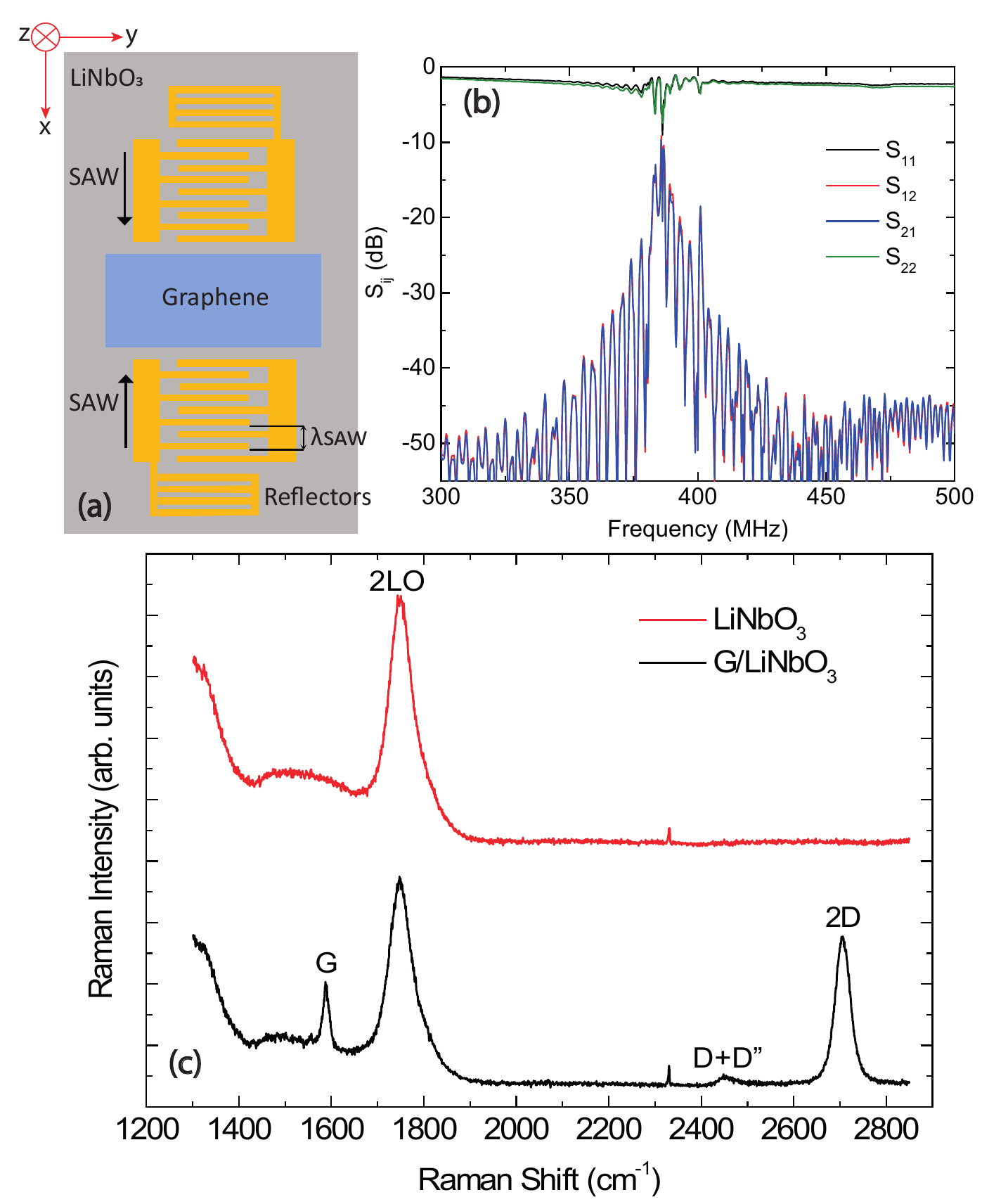}
\caption{(a) SAWR layout schematic. (b) S-parameters of the SAW device measured as a delay line. (c) Raman spectra of the bare LiNbO$_{\text{3}}$ substrate and the graphene/LiNbO$_{\text{3}}$ system.}
\end{figure}

\section{Results and Discussion}

\begin{figure*}
\includegraphics[width=\textwidth]{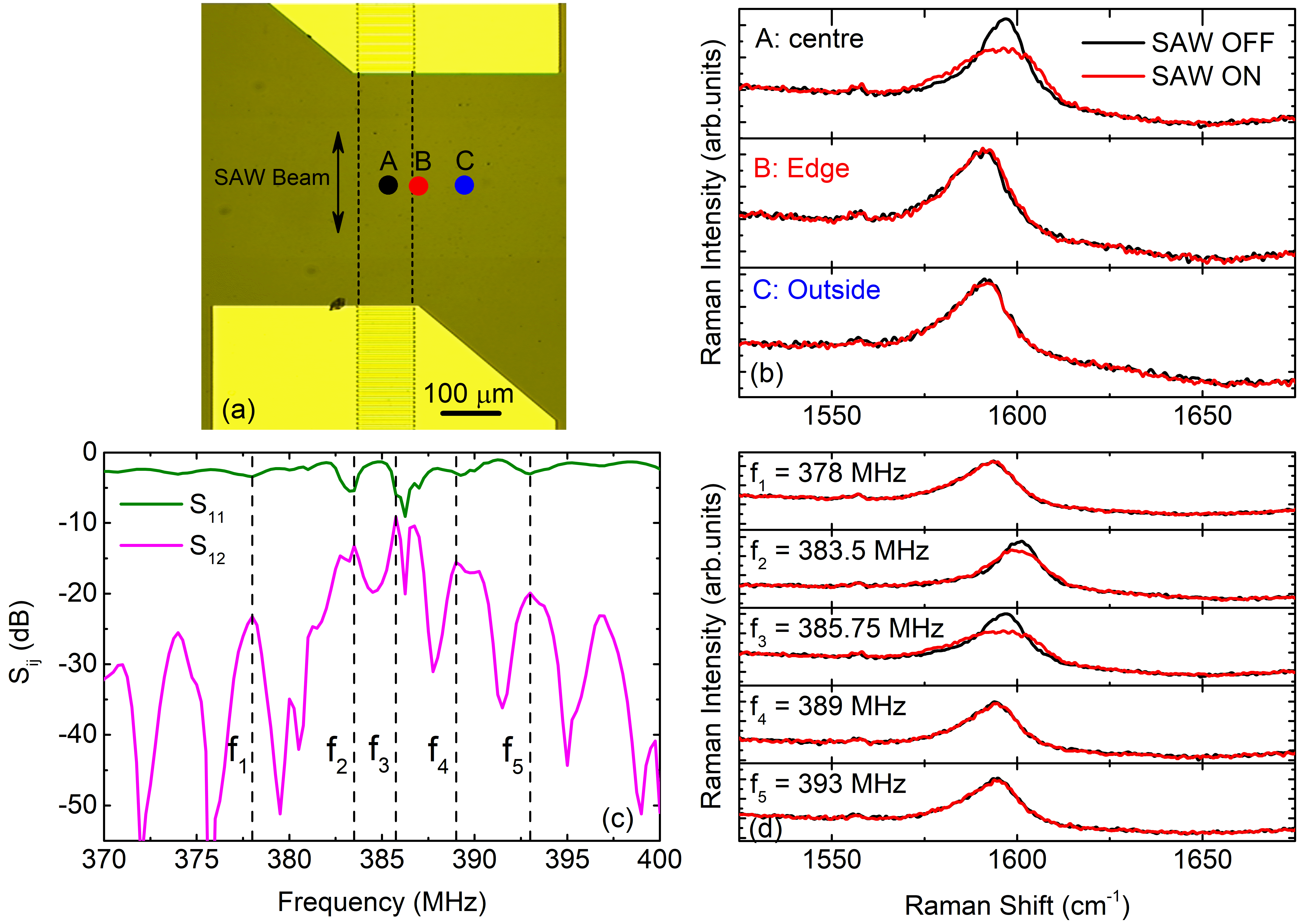}
\caption{(a) Optical micrograph of the SAWR with the transferred graphene in between the IDTs.(b) Raman spectra of the G peak for the \textit{SAW ON} and \textit{SAW OFF} conditions ($ f= 385.75 $ MHz and $ P_{RF}= 23.69 $ dBm) at three different positions on the device marked as A, B, and C in (a). (c) $S_{11}$ (reflection) and $S_{12}$ (transmission) losses of the device. (d) Raman spectra of the G peak for the \textit{SAW ON} and \textit{SAW OFF} conditions ($ P_{RF}= 23.69 $ dBm) at the different frequencies indicated by dashed vertical lines in (c).}
\end{figure*}

The Raman spectra of the bare LiNbO$_{\text{3}}$ substrate and of the SLG transferred onto it are shown in figure 1(c). LiNbO$_{\text{3}}$ exhibits a strong Raman peak around 1780 cm$^{-1}$ which is a two longitudinal optical phonon process \cite{Kokanyan15}. The characteristic Raman signature of the SLG, formed by the G and 2D peaks at 1580 and 2700 cm$^{-1}$ respectively, appears on top of the background provided by the LiNbO$_{\text{3}}$ substrate. A small D+D'' peak is also observed at 2425 cm$^{-1}$. The intensity ratio $I(2D)/I(G)\approx 2.5 $ and the absence of the characteristic defect (D) peak (at $\sim$ 1350 cm$^{-1}$) confirms that the graphene is a defect free monolayer. The analysis of the frequency correlation for the G and 2D peaks indicates that the SLG has a residual carrier density of approximately $ 4.5\:$ x $\: 10^{12}$ cm$^{-2}$ and a native strain, arising from the transfer process and the lattice mismatch with the substrate, of $-0.1\:\%$, where the minus sign indicates compressive strain (see appendix A for details).

\begin{figure}
\includegraphics[width=\linewidth]{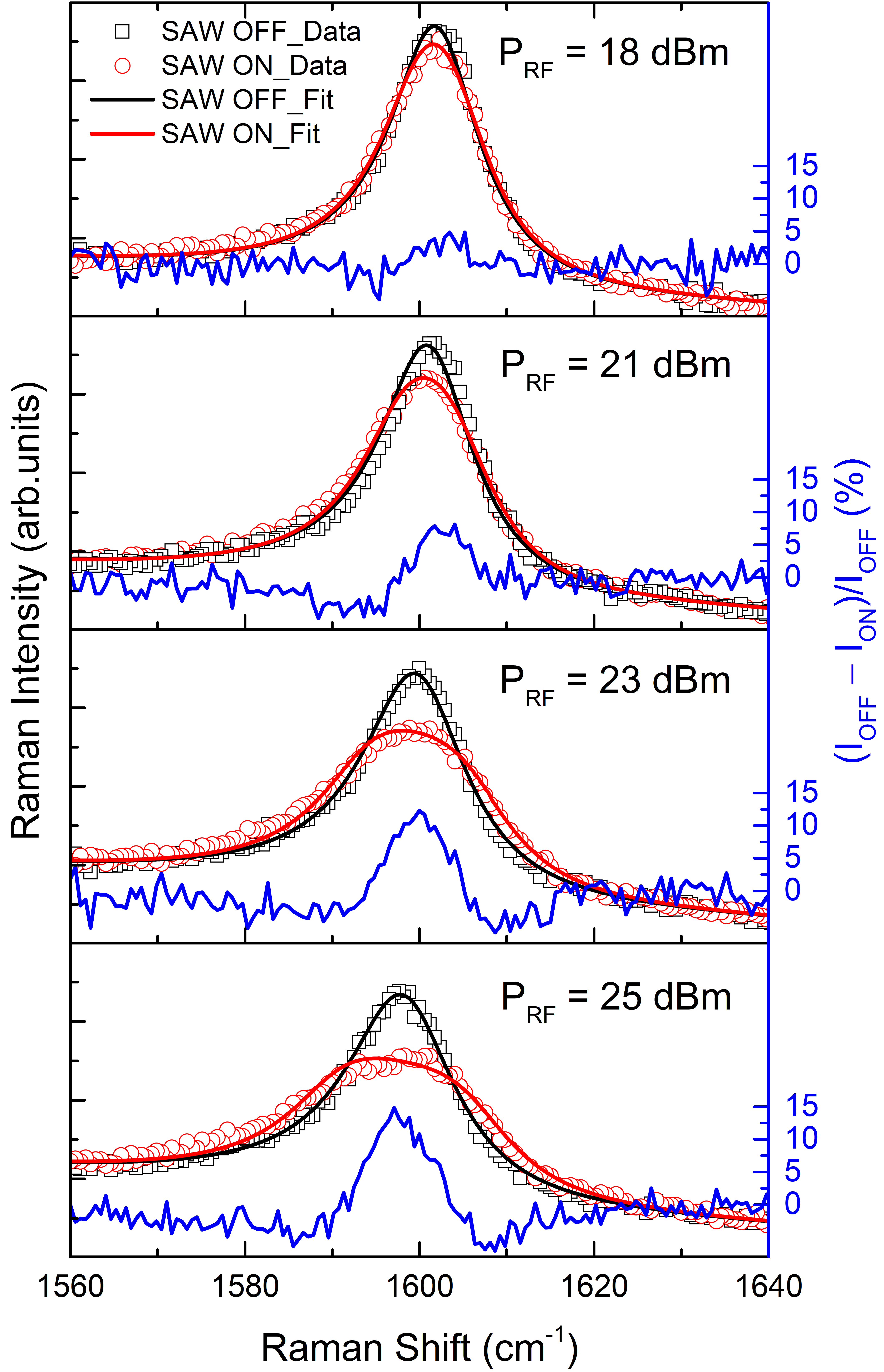}
\caption{Raman spectra of the G peak for the \textit{SAW OFF}, $ I_{off} $, and \textit{SAW ON}, $ I_{on} $, conditions for $ f= 385.75 $ MHz and increasing values of the applied RF power $ P_{RF} $. The experimental spectra (symbols) are fitted (lines) with Eq.~(\ref{eq:1}) and Eq.~(\ref{eq:2}), respectively. The net differential response $ (I_{off} - I_{on})/I_{off} $ is also depicted (blue curve).}
\end{figure}

\begin{figure}
\includegraphics[width=\linewidth]{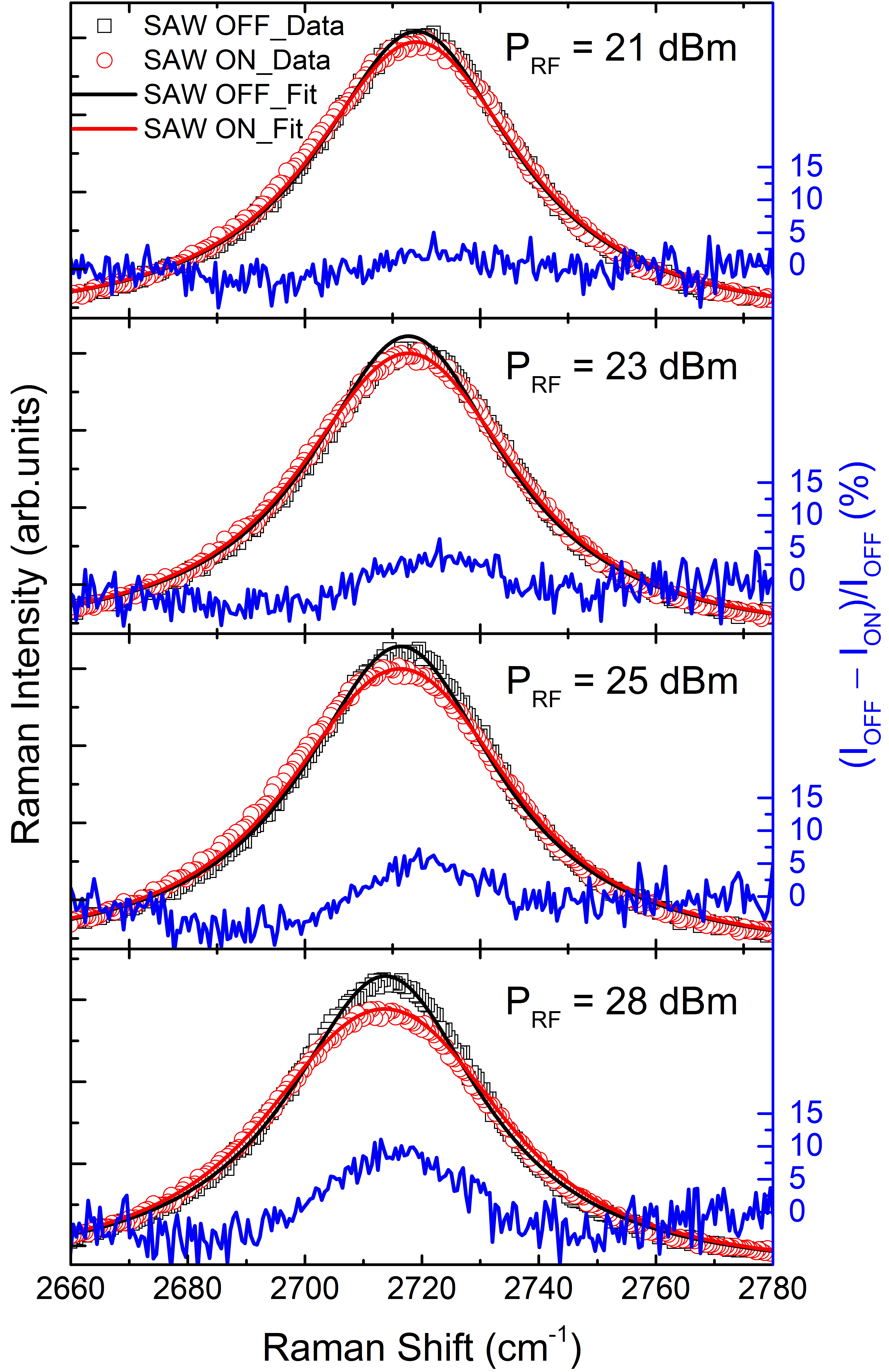}
\caption{Raman spectra of the 2D peak for the \textit{SAW OFF}, $ I_{off} $, and \textit{SAW ON} conditions, $ I_{on} $, for $ f= 385.75 $ MHz and increasing values of the applied RF power $ P_{RF} $. The experimental spectra (symbols) are fitted (lines) with Eq.~(\ref{eq:1}) and Eq.~(\ref{eq:2}), respectively. The net differential response $ (I_{off} - I_{on})/I_{off} $ is also depicted (blue curve). Larger $ P_{RF} $ values were applied here, as compared to the G peak in figure 3, to achieve a noticeable modulation.}
\end{figure}

We have investigated the effects of the SAW on the G and 2D Raman bands of the SLG. These bands correspond to the twofold degenerate $E_{2g}$ optical phonons at the $\Gamma$ point of the Brillouin zone (in-plane C-C stretching mode) and the two $A_{1g}$ optical phonon scattering at the K (and K') point of the Brillouin zone (in-plane breathing-like mode), respectively. The SAWR produces two counter-propagating travelling SAWs forming a standing SAW pattern with a maximum amplitude at the anti-nodes that is twice as that of a single travelling SAW. The local effect of the SAWs has been first assessed by measuring the Raman spectra at different positions on the sample, marked as A, B, and C on the optical micrograph of the device shown in figure 2(a). The corresponding Raman spectra of the G peak for the \textit{SAW ON} and \textit{SAW OFF} cases are depicted in figure 2(b). The intensity modulation and broadening of the peak is only observed within the path of the SAW beam (position A), whereas the Raman spectrum remains unaltered at the edge of and outside that region (positions B and C, respectively). This proves that the modulation observed is solely due to the SAW strain field. In addition, the frequency of the RF signal applied to the IDTs has also been varied across the passband region of the SAWR, as shown in figure 2(c). The dips in $S_{11}$ (and peaks in $S_{12}$) are due to the different modes of the acoustic resonator. In this case, the maximum phonon modulation is confirmed to peak at a frequency of 385.75 MHz, figure 2(d), where the  transmission losses ($S_{12}$) of the SAW devices are minimum. This modulation fades away rapidly out of the resonance frequency and is negligible for the main side lobes (at 378, 389, and 393 MHz). This also proves that the modulation observed is solely driven by the SAW strain field and rules out any thermal contribution by the laser illumination \cite{Papasimakis2015} and the RF dissipation. Furthermore, it demonstrates that the graphene phonons can be modulated dynamically at high frequencies by tuning the RF signal at the IDT. The slight variations in the center wavenumber of the G peak observed in figure 2(b) at different laser positions on the sample are due to the variations in the native strain and residual doping of the graphene (as shown in appendix A), whereas those observed in figure 2(d) for the same laser position and different RF frequencies are produced by small mechanical and thermal drifts. However, these drifts do not affect the differential analysis of the \textit{SAW OFF} and \textit{SAW ON} conditions measured at a fixed RF frequency and power, as the above mentioned modulation method cancels out the effects of any systematic fluctuations on the Raman signal (as explained in the experimental section).

Figure 3 displays the Raman spectra of the G peak for various $ P_{RF} $ values for both \textit{SAW ON}, $ I_{on} $, and \textit{SAW OFF}, $ I_{off} $, cases. The net differential response $ (I_{off} - I_{on})/I_{off} $ is also depicted. The $ I_{off} $ spectra have an asymmetric line shape, which arises from the coupling of a discrete phonon spectrum to a continuum of electronic excitations. To account for this line shape, we have fitted the spectra with a Breit-Wigner-Fano (BWF) function \cite{Hasdeo2014}, so that $ I_{off} $ reads as follows:
\begin{equation}\label{eq:1}
I_{off}(\nu)=I_{0}+b\nu+\frac{H\left(1+\frac{\nu-\nu_{c}}{q_{BWF}\Gamma}\right)^{2}}{1+\left(\frac{\nu-\nu_{c}}{\Gamma}\right)^{2}},
\end{equation}
where $I_{0}$ is the base line, $ b $ is the slope of straight line used for background removal, $ H $ is the height of the peak, $\nu_{c}$ is the centre of the peak, $ \Gamma $ is the spectral width, and $ 1/q_{BWF} $ is the asymmetry factor. The condition $1/q_{BWF}=0$ cancels out the contribution from the continuum of electronic excitations, giving a Lorentzian line shape, which represents a discrete phonon spectrum. The corresponding $ I_{on} $ Raman spectra show an additional line broadening of the G peak and a decrease in $ H $ for a certain range of $ P_{RF} $ values. This broadening in the G peak line shape is interpreted to be caused by the dynamic strain field of the SAW. Thus, the oscillating tensile and compressive cycles of the strain modulate periodically the phonon frequency around its equilibrium value. Since the acquisition time of the Raman spectra is much larger than the SAW period, the phonon oscillation manifests itself as a broadening of the peak line shape and a reduction of its maximum intensity. Both the broadening and the intensity reduction increase with increasing $ P_{RF} $.
 
The effects of the local SAW strain observed in the graphene G and 2D Raman lines are found to be qualitatively similar to those described in the modulation of the LO phonon Raman line of bulk semiconductor crystals \cite{Iikawa2016}. They are also similar to those observed in the photoluminescence of SAW-modulated quantum wells \cite{Sogawa2001} and dots \cite{Gell2008,Pustiowski2015}.  
In Si, the differential Raman signal is slightly asymmetric \cite{Iikawa2016}. The asymmetry was attributed to the scattering events under a SAW involving LO phonons with small wave vectors, which have energies slightly lower than the one for the zone-center mode. This type of asymmetry has not been observed for the graphene G Raman line, thus indicating that these scattering events are not important. Furthermore, the piezoelectric field accompanying the SAW can modulate the carrier concentration in semiconductors leading to broadenings and shifts of the Raman frequency \cite{Cerdeira1972}. In the case of SLG, however, the SAW-induced carrier density modulation has been estimated to be of the order of $10^{10}$ cm$^{-2}$, which is negligible as compared to the residual carrier density in the material.
Nonetheless, the SAW strain modulation observed in graphene is quantitatively much stronger as in the semiconductors due to the combination of the larger Grüneisen parameters of graphene and the more efficient generation of the strain in the 2D layer on a strong piezoelectric substrate, as we will discuss in the next paragraphs.  

\begin{figure}
\includegraphics[width=\linewidth]{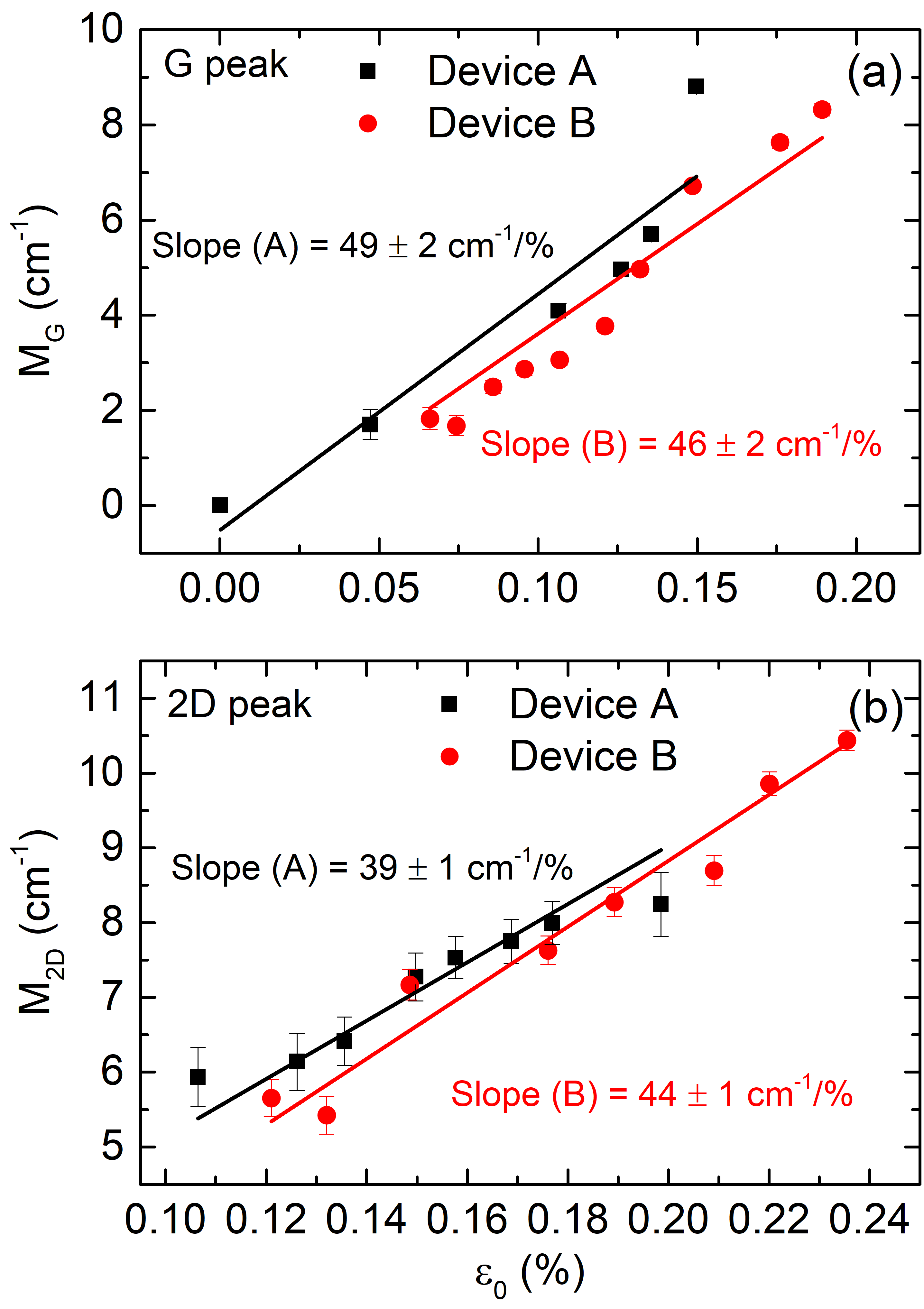}
\caption{Absolute value of the modulation, $ M_{G(2D)} $, of the (a) G- and (b) 2D-peak frequency as a function of the absolute value of the hydrostatic strain, $\varepsilon_{0}$, produced by the SAW for devices A and B using graphene from two different batches. Experimental values (symbols) and their linear fittings (lines) are shown. The slight differences in the behaviour of the two devices are attributed to variations in the graphene quality and the IDT efficiency.}
\end{figure}

In order to quantify the modulation of the graphene phonon by the strain field of the standing SAW, we have fitted the $ I_{on} $ spectra with the following integral equation:
\begin{equation}\label{eq:2}
I_{on}(\nu)=I_{0}+b\nu+\frac{1}{2\pi} \int_{0}^{2\pi}\frac{H\left(1+\frac{\nu-\left(\nu_{c}-M\cos{\theta}\right)}{q_{BWF}\Gamma}\right)^{2}}{1+\left(\frac{\nu-\left(\nu_{c}-M\cos{\theta}\right)}{\Gamma}\right)^{2}}d\theta,
\end{equation}
where $I_{0}$, $ b $, $ H $, $\nu_{c}$, $ \Gamma $, and $ 1/q_{BWF} $ are the fitting parameters extracted previously from the BWF fitting of $I_{off}(\nu)$ for each value of $ P_{RF} $. The amplitude of the SAW-induced phonon frequency modulation is given by $ M $, which is the only free parameter in the fitting of $ I_{on}(\nu) $. $ M $ is the absolute value of the SAW-induced modulation, whereas its sign variation within the wave cycle is accounted for by the cosine term. $ M $ attains a non-zero value only within the SAW beam path, for the SAW resonance frequency, and for a large enough RF power. We have also performed a similar study for the 2D peak, as shown in figure 4. It must be noted that in the case of the 2D peak, $ 1/q_{BWF} $ is very small, as expected for a band hardly affected by the continuum of electronic excitations. A weaker acoustic modulation is observed for the 2D band as compared to the G band, so that much higher $ P_{RF} $ values have been required to obtain a comparable modulation to that of the G band. This is expected to be related to the larger width of the 2D peak that partially hinders the modulation making it less apparent. 

Figures 5(a) and 5(b) display the evolution of the value of $ M $ for the G, $ M_{G} $, and 2D, $ M_{2D} $, peaks, respectively, as a function of the absolute value of the hydrostatic strain, $ \varepsilon_{0}$, generated by the standing SAW at the surface of the LiNbO$_{3}$ substrate where the SLG lies. Since the SAW is a Rayleigh mode \cite{Rayleigh1885}, i.e. with sagittal polarization, it has only particle displacements along the x and z directions (cf. figure 1(a)). Therefore, the hydrostatic component of the strain field is then $\varepsilon_{0}=\varepsilon_{xx}+\varepsilon_{zz}$. The details of the calculation of $\varepsilon_{0}$ as a function of the $ P_{RF} $ applied to the IDTs are described in appendix B. Figures 5(a) and 5(b) display the data from two devices (A and B) with identical SAWR but with transferred graphene from two different batches. The linear fittings in figure 5(a) provide a slope value or rate $\partial M_{G}/\partial\varepsilon_{0}$ of 49 $ \pm $ 2 and 46 $ \pm $ 2 cm$^{-1}/\%$ for devices A and B, respectively. Similar fittings in figure 5(b) provide a rate $\partial M_{2D}/\partial\varepsilon_{0}$ of 39 $ \pm $ 1 and 44 $ \pm $ 1 cm$^{-1}/\%$, respectively, for the same devices. The obtained values are then consistent for both graphene batches used. It has to be noticed that although the strain produced by the SAWR has a quasi-uniaxial character, since $ \varepsilon_{zz} $ is an order of magnitude smaller than $ \varepsilon_{xx} $, the polycrystalline nature of the CVD SLG used for this study averages out the contribution from several crystal domains, so that the well-known G- \cite{Huang2009,Mohiuddin2009,Frank2010} and 2D-peak \cite{Huang2010,Frank2011,Yoon2011} splitting in (monocrystalline) exfoliated graphene flakes under uniaxial strain is not expected to occur here. Nevertheless, these splittings are foreseen to be induced by the SAWR in SLG monocrystals (CVD or exfoliated).

The $\partial M_{G(2D)}/\partial\varepsilon_{0}$ rates obtained have allowed us to calculate the respective Grüneisen parameters, $ \gamma_{G(2D)} $, which represent the rate of phonon mode softening (hardening) under tensile (compressive) strain. They are expressed as \cite{Note1} 
\begin{equation}\label{eq:3}
\gamma_{G(2D)}=\dfrac{1}{\nu^{G(2D)}_{c}}\dfrac{\partial M_{G(2D)}}{\partial\varepsilon_{0}},
\end{equation}
where $ \nu^{G(2D)}_{c} $ is the centre of the position of the G (2D) peak for the unperturbed (\textit{SAW OFF}) condition. The values of $ \gamma_{G} $ obtained for devices A and B are 3.1 $ \pm $ 0.1 and 2.8 $ \pm $ 0.1, respectively, whereas the values of the corresponding $ \gamma_{2D} $ are 1.43 $ \pm $ 0.04 and 1.63 $ \pm $ 0.04, respectively. 

The Grüneisen parameters are fundamental magnitudes that determine the thermomechanical properties of a material. However, in the case of graphene, there is still a large dispersion of values reported in the literature, with $ \gamma_{G} $ and $ \gamma_{2D} $ in the ranges of $ 0.69-2.4 $ and $ 2.98-3.8 $, respectively \cite{Huang2009,Mohiuddin2009,Ding2010,Metzger2010,Cheng2011}. This variation can be attributed to the different type of strain applied, as well as to the quality of the graphene, the substrate used \cite{Jiang2018}, and the effective adhesion of the graphene to the substrate \cite{Bousige2016}. The smaller value of $ \gamma_{2D} $ than of $ \gamma_{G} $ obtained in this study seems to be an artifact related to the less efficient SAW-induced modulation of the 2D band due to its broader nature, as it has been discussed in figure 4. On the other hand, the larger values of $ \gamma_{G} $ obtained in this study, as compared to those reported in the literature, might arise from the large difference in the coefficient of thermal expansion, $ CTE $, of the polymers used in the case of graphene on flexible substrates, such as SU8 in ref. \cite{Mohiuddin2009} ($ CTE = 5.2$ x $10^{-5}$ K$^{-1}$ \cite{Steiner2012}) or PDMS in ref \cite{Huang2009} ($ CTE = 9-9.6$ x $10^{-4}$ K$^{-1}$ \cite{Mark2009}), and of LiNbO$_{\text{3}}$ ($CTE = 7.5-15$ x $10^{-6}$ K$^{-1}$ \cite{Kim1969}), as well as due to the dynamic nature of the strain produced by the SAW on the graphene/LiNbO$_{\text{3}}$ system in contrast to the static strain in the case of the graphene/polymer systems on a flexible substrate.

Most bulk semiconductors have Grüneisen parameters for the zone-center TO and LO (or degenerated LTO) phonons $ \gamma_{LTO} $ with values around 1 \cite{Yu2010} and phonon frequencies $ \nu^{LTO}_{c} $ in the range of 250-550 cm$^{-1}$. Therefore, the Raman frequency sensitivity to strain of these materials is expected to be of approximately 5 cm$^{-1}/\%$, an order of magnitude smaller than the $\partial M_{G(2D)}/\partial\varepsilon_{0}$ rates obtained here for graphene modulated by the SAW strain field. In addition, graphene can be easily transferred directly onto a strong piezoelectric substrate providing a large SAW strain amplitude. Conversely, semiconductors are typically weakly or non-piezoelectric, the latter requiring an additional piezoelectric layer on top of them to generate the SAW. Thus, in these cases, the SAW strain amplitude is either small or only a fraction of that produced at the piezoelectric layer reaches the active semiconductor region \cite{Pedros2011}. This makes SAWs specially powerful and well-suited tools for the strain modulation of graphene and the development of novel devices thereof.

\section{Conclusions}
In conclusion, we have investigated the SAW-induced modulation of the phonons of graphene by Raman spectroscopy. We have observed strong effects in both first-order and second-order (or two-phonon) Raman modes, achieving intensity variations of $15\:\%$ and phonon frequency shifts of around $10$ cm$^{-1}$ when the SAW is applied. The local strain produced by the SAW can be varied by changing the applied RF power and frequency allowing us to dynamically tune the phonon frequency. SAWs might also permit to generate strain along multiple axes and tailor the strain fields. For example, a biaxial strain pattern could be generated by either using orthogonal SAWRs or selecting crystal cuts and propagation directions providing comparable $ \varepsilon_{xx} $ and $ \varepsilon_{yy} $ components \cite{Campbell1968}. An adequate design of the  in- and out-of-plane strain components of the SAW might also be useful for tailoring gauge fields in graphene \cite{Guinea2009, Vozmediano2010}. Moreover, the amount of strain generated in the SLG could be enhanced by using focusing IDTs. Furthermore, the piezoelectric field accompanying the SAW in LiNbO$_{\text{3}}$ leads to a variation of the SLG carrier density of the order of $ 10^{10}$ cm$^{-2} $, which could be used for locally modulating the material near the Dirac point.
 
The modulation mechanism reported here is not limited to single-layer graphene but can be extended to any other single- or few-layer 2D material, where the physics is specially rich and strain engineering opens a whole range of new possibilities \cite{Amorim2016,Naumis2017}. Moreover, many 2D materials are themselves piezoelectric \cite{Cui2018}, enhancing the capabilities of strain modulation. As an example, the $E_{2g}$ mode has been shown to strongly modify the plasmonic properties of bilayer graphene \cite{Low2014}, so that a SAW could be used to tune hybridized plamon-phonon modes for surface-enhanced Raman spectroscopy \cite{Marin2017}.

\begin{acknowledgments} 
The authors thank Manfred Ramsteiner for helpful discussions and Adolfo del Campo (ICV-CSIC) for his assistance with the preliminary Raman measurements in appendix A. This work has received funding from the European Union’s Horizon 2020 Research and Innovation Programme under Marie Skłodowska-Curie Grant Agreement No 642688, from the Spanish Ministry of Economy and Competitiveness (MINECO) through project DIGRAFEN (ENE2017-88065-C2-1-R), from the Comunidad de Madrid through project NMAT2D-CM (P2018/NMT-4511), and from the Universidad Politécnica de Madrid through project GRAPOL (VJIDOCUPM18JPA). J.P. acknowledges financial support from MINECO (Grant RyC-2015-18968). F.I. acknowledges the Alexander von Humboldt Foundation and the Conselho Nacional de Desenvolvimento Científico e Tecnológico (Nos. 305769/2015-4 and 432882/2018-9) for financial support. 
\end{acknowledgments}

\appendix

\section{Graphene native strain} 

\begin{figure} [hbt]
\includegraphics[width=\linewidth]{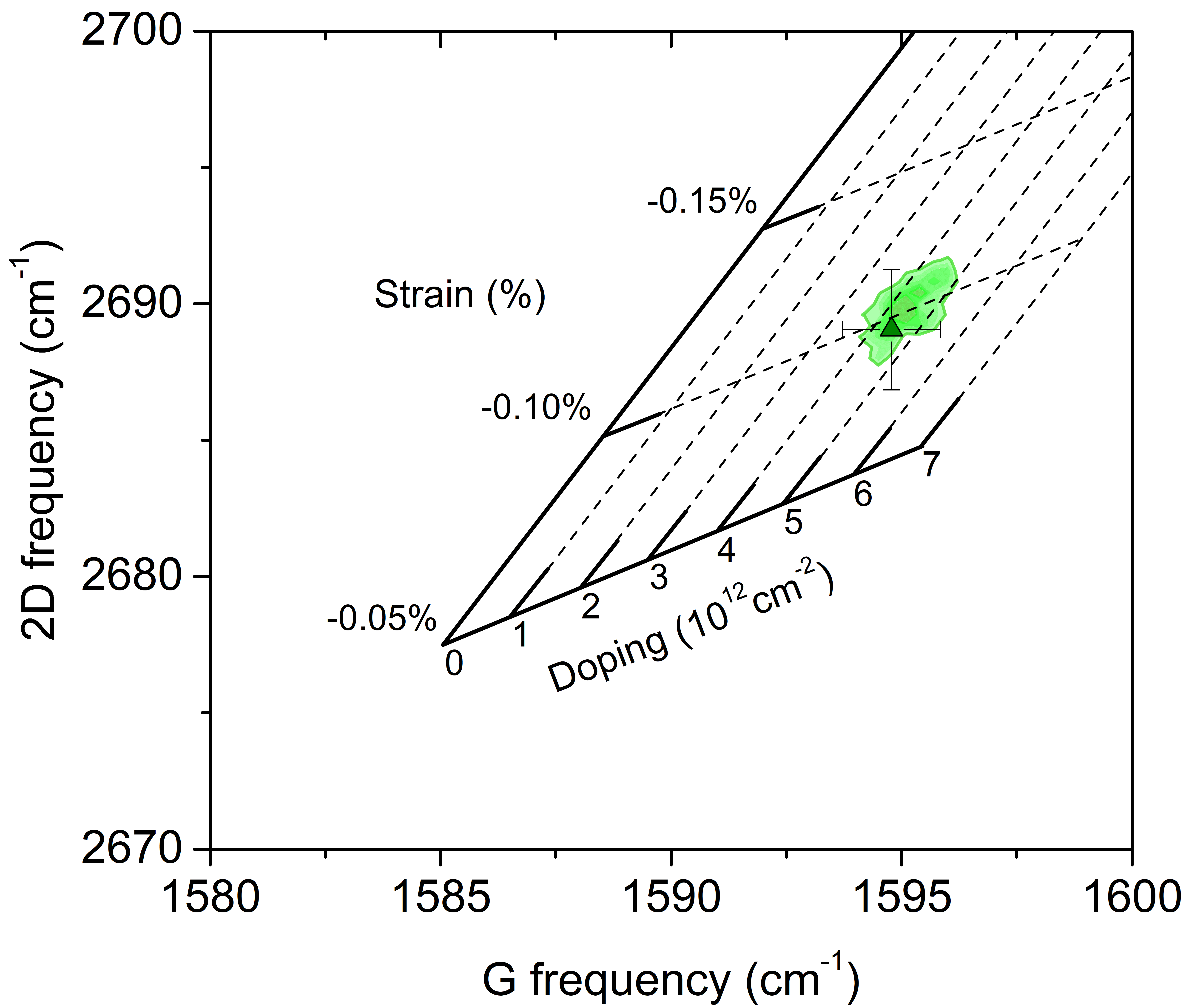}
\caption{Correlation of the G- and 2D-peak frequencies of the graphene transferred to LiNbO$_{\text{3}}$. The mean and standard deviation of the distribution is shown by the symbol and the error bars, respectively. The embedded internal axes decouple the doping and the strain effects.}
\end{figure}

A WITec Alpha 300AR Raman confocal microscope was used for the preliminary Raman spectroscopy mapping of the samples. Raman spectra were obtained in backscattering geometry using a 50x objective lens (numerical aperture of 0.8) in ambient conditions. A 532-nm wavelength laser
set to a power of 1 mW was used as excitation source. Typical mapped areas were 80 x 100 $ \mu $m in size, scanned with steps of 4 $ \mu $m. 

Figure 6 shows the correlation between the frequencies of the G and 2D peaks in one of these Raman mappings. The larger excitation wavelength used here as compared to the one used for the Raman spectroscopy under a SAW leads to smaller frequencies of the 2D peak \cite{Zhao2011} than in previous figures. The diagonal evolution of the data distribution shows a slope value of $ \partial\nu_{2D}/\partial\nu_{G} \approx 2.2 $, which is consistent with graphene under biaxial strain, $ (\partial\nu_{2D}/\partial\nu_{G})_{\varepsilon_{biaxial}} = 2.2 $  \cite{Metten2014,Shin2016} and constant p-type doping, $ (\partial\nu_{2D}/\partial\nu_{G})_{p} = 0.7 $  \cite{Lee2012b}. Using the reported sensitivity values of $\partial\nu_{G}/\partial\varepsilon_{biaxial}=-69.1 $  cm$^{-1}/\% $ and $\partial\nu_{2D}/\partial p=-1.04 $  cm$^{-1}/10^{12}$ cm$^{-2} $ \cite{Lee2012b}, these frequencies are decomposed in a native strain and residual doping-related basis \cite{Lee2012b}, as shown by the internal set of axes in figure 6.

\section{SAW-induced strain vs power density}

\begin{figure} [hbt]
\includegraphics[width=\linewidth]{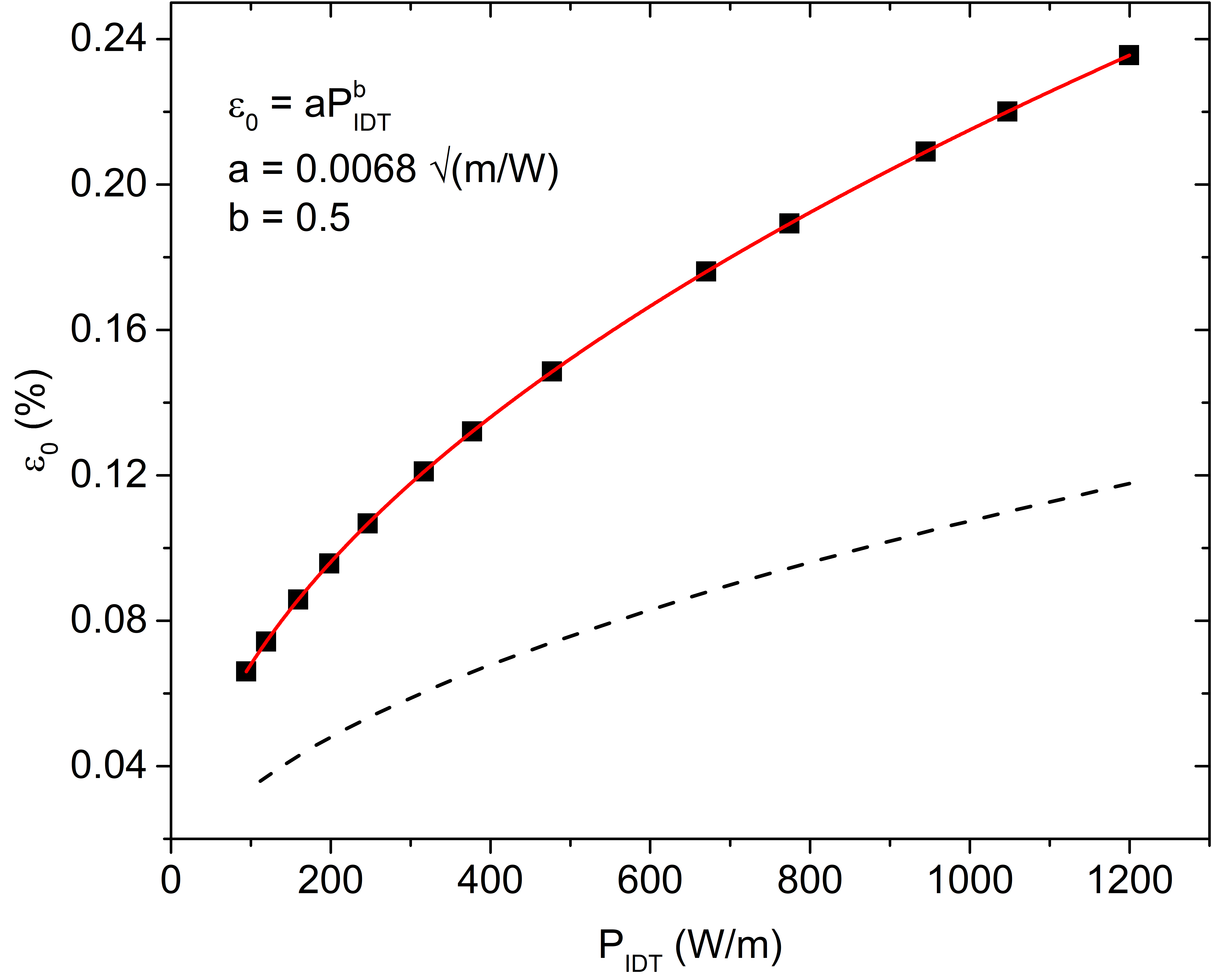}
\caption{Amplitude of the hydrostatic strain $\varepsilon_{0}$ produced by the SAWR at the anti-nodes of the standing wave pattern (both IDTs excited, solid line) as function of the SAW power density generated by each IDT $ P_{IDT} $ (W/m). For comparison, the strain associated to a travelling wave generated by a single IDT (dashed line) is also depicted.}
\end{figure}

The output signal from an RF generator was split and amplified, in order to drive each IDT of the SAWR with an RF power $ P_{RF} $ (dBm). The SAW power density (power per unit of SAW beam width) generated by each IDT was then calculated using the following relation
\begin{equation}\label{eq:4}
P_{IDT}(W/m)=\dfrac{10^{[\dfrac{(S_{12}/2)(dB)}{10}]}10^{[\dfrac{(P_{RF}(dBm)-30)}{10}]}}{W_{IDT}(m)}.
\end{equation}
The half of the transmission parameter $ S_{12}(dB) $ of the SAWR takes into account the electromechanical efficiency of a single IDT and reflection grating.

Rayleigh SAWs have a mixed compressional and shear character, leading to a displacement field vector 
\begin{equation}\label{eq:5}
u_{R}=(u_{x}, 0, u_{z}),
\end{equation}
where $ u_{x} $ and $ u_{z} $ are the longitudinal and shear vertical components, respectively, and the shear horizontal component $ u_{y} $ is typically zero. In the particular case of $128^{o}$ rotated Y-cut X-propagating LiNbO$_{\text{3}}$, $ u_{y} $ is not zero but approximately 20 times smaller than the other two components, so it has been neglected.
Thus, using the Voigt notation, the strain field is written as 
\begin{equation}\label{eq:6}
\varepsilon_{R}=(\varepsilon_{xx}, 0, \varepsilon_{zz}, 0, 2\varepsilon_{xz}, 0),
\end{equation}
where the strain tensor components are given by
\begin{equation}\label{eq:7}
\varepsilon_{kl}=\dfrac{1}{2}(\dfrac{\partial u_{k}}{\partial x_{l}}+\dfrac{\partial u_{l}}{\partial x_{k}})
\end{equation}
and the zero components in equation (B3) appear because $ u_{y}=\partial u_{x}/\partial y=\partial u_{z}/\partial y=0 $.
The particle displacements have been calculated by solving numerically the coupled elastic and electromagnetic equations for the LiNbO$_{\text{3}}$ substrate. This allow us to calculate the strain components and the hydrostatic strain $ \varepsilon_{0}=\varepsilon_{xx}+\varepsilon_{zz} $ produced by the SAWR in the LiNbO$_{\text{3}}$ substrate, and transmitted to the graphene layer on top.  

Figure 7 shows the amplitude of $\varepsilon_{0}$ as a function of $ P_{IDT} $ (W/m), where a square root dependence is observed. The maximum strain is achieved at the anti-nodes of the standing wave produced by the SAWR (where both IDTs are excited), which is 2 times larger than that generated by a travelling wave excited by a single IDT. The sign of the strain depends in both cases on the part of the cycle of the wave.

\nocite{*}
\bibliographystyle{iopart-num}
\bibliography{ref}

\end{document}